\documentstyle[12pt,epsfig]{article}
\newcommand{\Frac}[2]%
{{\textstyle \frac{\mbox{\footnotesize $#1$}\rule[-0.9mm]{0mm}{1mm}}%
{\mbox{\footnotesize $#2$}\rule{0mm}{3.1mm}}}}
\renewcommand{\thefootnote}{\fnsymbol{footnote}}

\begin{document}

\begin{titlepage}
\vspace*{-12 mm}
\noindent
\begin{flushright}
\begin{tabular}{l@{}}
\end{tabular}
\end{flushright}
\vskip 12 mm
\begin{center}
{\large \bf
The cosmological constant puzzle: 
%%\\
Vacuum energies from QCD to dark energy
}
\\[14 mm]
{\bf Steven D. Bass}
\\[5mm]
{\em
Institute for Theoretical Physics, \\
Universit\"at Innsbruck,
Technikerstrasse 25, Innsbruck, A 6020 Austria
\\[10mm]
}
\vspace{0.5cm}
\end{center}
\vskip 20 mm
\begin{abstract}
\noindent 
The accelerating expansion of the Universe points to a small
positive vacuum energy density and negative vacuum pressure.
A strong candidate is the cosmological constant in Einstein's
equations of General Relativity.
The vacuum dark energy density extracted from astrophysics is 
10$^{56}$ times smaller than the value expected from 
the Higgs potential in Standard Model particle physics.
The dark energy scale is however close to the range of possible values expected for the light neutrino mass. 
We investigate this physics in a simple toy model
where the chirality of the neutrino is treated 
by analogy as an Ising-like ``spin" degree of freedom.

\end{abstract}

\vspace{9.0cm}

\end{titlepage}
\renewcommand{\labelenumi}{(\alph{enumi})}
\renewcommand{\labelenumii}{(\roman{enumii})}
\renewcommand{\thefootnote}{\arabic{footnote}}
\newpage
\baselineskip=6truemm

\section{Introduction}

The vacuum energy density perceived by gravitation drives accelerating expansion of the Universe.
Understanding this vacuum energy is an important challenge 
for theory and 
connects the Universe on cosmological 
scales (the very large) with subatomic physics (the very small);
for reviews see 
\cite{weinberg,astrid,bass:2010,Ruiz,Amendola}.

The physical world we observe today is built from spin-${1 \over 2}$ fermions interacting through the exchange of gauge bosons:
massless spin-1 photons and gluons;
massive W and Z bosons;
and gravitational interactions.
QED is manifest in the Coulomb phase,
QCD is manifest in the confinement phase and
the electroweak interaction is manifest in the Higgs phase.
Further ingredients are needed to allow the formation of large-scale
structures on the galactic scale and to explain the accelerating
expansion of the Universe. 
These are the mysterious dark matter and
dark energy, respectively.
Current observations point to an energy budget of the Universe where
just 4\% is composed of atoms, 23\% involves dark matter
(possibly made of new elementary particles) and 73\% 
is dark energy
(the energy density of the vacuum perceived by gravitational interactions).

The simplest explanation of this dark energy is a small positive value for the cosmological constant in Einstein's equations of General Relativity.
Einstein's equations link the geometry of spacetime to the 
energy-momentum tensor
\begin{equation}
R_{\mu \nu} - {1 \over 2} g_{\mu \nu} R
=
- {8 \pi G \over c^2} T_{\mu \nu} + \Lambda g_{\mu \nu} .
\end{equation}
Here $R_{\mu\nu}$ is the Ricci tensor which is built 
from the metric tensor $g_{\mu \nu}$ and its derivatives, 
$R$ is the Ricci scalar and 
$T_{\mu \nu}$ 
is the energy momentum tensor.
The left-hand side describes the geometry and 
the right-hand side describes 
the energy content of the gravitational system.
Writing
$
\Lambda = 8 \pi G \rho_{\rm vac} + \Lambda_0
$,
the cosmological constant tells us about the energy density of 
the vacuum $\rho_{\rm vac}$ perceived by gravitational interactions;
$\Lambda_0$ is a possible counterterm.

The vacuum energy density receives possible contributions from
the zero-point energies of quantum fields and condensates associated 
with spontaneous symmetry breaking.
The vacuum is associated with various condensates.
The QCD scale associated with quark and gluon confinement is around
1 GeV, while the electroweak mass scale associated with the W and Z boson masses is around 250 GeV.
These scales are many orders of magnitude less than the Planck-mass scale of around 10$^{19}$ GeV, 
where gravitational interactions are supposed to be sensitive to quantum effects.
If the net vacuum energy is finite it will have gravitational effect.
Being proportional to $g_{\mu \nu}$, a positive cosmological constant corresponds to negative pressure in the vacuum perceived by gravitational interactions.
The vacuum energy density associated with dark energy is characterised 
by a scale around 0.002 eV, typical of the range of possible 
light neutrino masses, and a cosmological constant, which is 56 orders 
of magnitude less than the value expected from the Higgs condensate with no extra new physics. 
Why is this vacuum ``dark energy" finite, and why so small?

The challenge presented by gravitation and the cosmological constant
is fundamentally different
from particle physics in that gravity couples
to everything whereas other physics processes and
experiments involve measuring the differences between quantities.

\section{Vacuum energy and the cosmological constant}

We next consider zero-point and condensate contributions to the vacuum energy.

Quantization introduces zero-point vacuum energies for quantum fields
and therefore, in principle, can affect the geometry through Einstein's equations.
Before normal ordering the zero-point energy of the vacuum is badly divergent, being the sum of zero-point energies for an infinite 
number of oscillators, one for each normal mode, or 
degree of freedom of the quantum fields \cite{bjd}.
Before interactions, the vacuum (or zero-point) energy is
\begin{equation}
\rho_{\rm vac} =
{1 \over 2} \sum \{\hbar\omega \}
=
{1 \over 2} \hbar
\sum_{\rm particles} g_i \int_0^{k_{\rm max}}
{d^3 k \over (2 \pi)^3} \sqrt{k^2 + m^2}
\sim
\sum_i {g_i k^4_{\rm max} \over 16 \pi^2} .
\end{equation}
Here ${1 \over 2} \{ \hbar \omega \}$ 
denotes the eigenvalues of the free Hamiltonian and
$\omega = \sqrt{k^2 + m^2}$
where $k$ is the wavenumber and $m$ is the particle mass;
$g_i = (-1)^{2j} (2j+1)$
is the degeneracy factor for a particle $i$ of spin $j$, 
with $g_i  >0$ for bosons and $g_i < 0$ for fermions.
The minus sign follows from the Pauli exclusion principle and 
the anti-commutator relations for fermions.
The vacuum energy density $\rho_{\rm vac}$ is quartically divergent 
in $k_{\rm max}$.

What value should one take for $k_{\rm max}$ ?

Possible candidates are the energy-scale associated with electroweak symmetry breaking 
$\Lambda_{\rm ew} 
= 2^{-1/4}G_F^{-1/2} = 246 \ {\rm GeV}$
and the Planck scale
$M_{\rm Pl} 
= \sqrt{ \hbar c / G } = 1.2 \times 10^{19} \ {\rm GeV}$
where we expect quantum gravity effects to become important.
Substituting $k_{\rm max} \sim \Lambda_{\rm ew}$ into Eq.~(2) 
with no additional physics gives a cosmological constant
\begin{equation}
\Lambda_{\rm vac} \sim 8 \pi G \Lambda_{\rm ew}^4
\end{equation}
or
\begin{equation}
\rho_{\rm vac} =
{1 \over 2} \sum \hbar \omega \sim (250 \ {\rm GeV})^4 .
\end{equation}
This number is 56 orders of magnitude larger than the observed value
\begin{equation}
\rho_{\rm vac} \sim (0.002 \ {\rm eV})^4 .
\end{equation}
Also, summing over just the Standard Model fields in Eq.~(2) gives a negative overall sign whereas the value of $\rho_{\rm vac}$ extracted from cosmology is positive.
What dilutes the large particle physics number to the physical value measured in large scale astrophysics and cosmology?
If we take $k_{\rm max} \sim M_{\rm Pl}$, then we obtain a 
value for $\rho_{\rm vac}$ which is $10^{120}$ times too big.

In quantum field theory (without coupling to gravity) the zero-point energy is removed by normal ordering so that the zero of energy is defined as the energy of the vacuum.
This can be done because absolute energies here are not measurable observables. 
Only energy differences have physical meaning, 
{\it e.g.} in Casimir processes~\cite{astrid,jaffe}, 
before we couple the theory to gravity.

Suppose we can argue away quantum zero-point contributions 
to the vacuum energy.
One still has to worry about spontaneous symmetry breaking.
Condensates that carry energy appear at various energy scales 
in the Standard Model,
{\it e.g.}
the Higgs condensate gives 
$
\rho_{\rm vac} \sim - (250 \ {\rm GeV})^4$
with negative sign.
The QCD condensate gives about $- (200 \ {\rm MeV})^4$.
These condensates form at different times in 
the early Universe,
suggesting some time dependence to $\rho_{\rm vac}$.
If there is a potential in the vacuum it will, in general, 
correspond to some finite vacuum energy.
Why should the sum of many big numbers 
(plus any possible gravitational counterterm) 
add up to a very small number?

\section{Seeking a possible explanation}

It is interesting that the dark energy or cosmological constant scale in Eq.(5) is of the same order of magnitude that we expect for the light neutrino mass, {\it viz.} 
$0.002 \ {\rm eV}$~\cite{altarelli,wetterich,nucond}
\begin{equation}
\mu_{\rm vac} \sim m_{\nu} \sim \Lambda_{\rm ew}^2/M .
\end{equation}
where $M \sim 3 \times 10^{16}$ GeV is logarithmically close 
to the Planck mass $M_{\rm Pl}$ and typical of the scale that 
appears in Grand Unified Theories.
Further, the gauge bosons in the Standard Model which have a mass through the Higgs mechanism are also the gauge bosons which couple 
to the neutrino. Is this a clue?
The non-perturbative structure of chiral gauge theories is not well
understood.~\footnote{We note previous investigations of the close value of the neutrino mass and the cosmological constant scale~\cite{wetterich,nucond}.
Ideas include time varying scalar fields with possible coupling 
to neutrinos (with corresponding varying mass) \cite{wetterich}
as well as possible neutrino condensates \cite{nucond}.
Neutrino condensates could be generated by introducing a 
new attractive 4-neutrino interaction into a 
BCS or Nambu-Jona-Lasinio like model,
induced by a new scalar or extra new physics since Z$^0$ exchange yields a repulsive vector interaction between left-handed neutrinos.
In these models one needs also additional new physics to 
remove the Higgs and QCD contributions to the net vacuum energy
and to worry about possible fine tuning issues associated with
couplings of the scalar field.
Possible time dependence of the fundamental parameters in 
particle physics induced by time dependent dark energy is discussed 
in \cite{hfsola}.
}

Changing the external parameters of the theory can change the phase
of the ground state.
For example, QED in 3+1 dimensions with exactly massless electrons is believed to dynamically generate a photon mass \cite{gribov}.
In the Schwinger Model for 1+1 dimensional QED on a circle, setting
the electron mass to zero shifts the theory from a confining
to a Higgs phase \cite{gross}.
In 1+1 dimensions the same result holds for SU(N)
where all the dynamical fields are in the adjoint representation and play a physical role similar to that of transverse gluons in 3+1 dimensional theories plus massless adjoint Majorana fermions 
\cite{gross}. 
Confinement  gives way to the screening of fundamental test charges and Higgs phenomena if the fermion mass is set exactly to zero. 
Gross et al. \cite{gross} write 
{\it
``The pure 4D Yang-Mills theory is expected to be confining.
In view of what we learned from 1+1 dimensional examples we 
may wonder, however, whether instead it could be in the screening phase: certain gluonic excitations might be capable of screening fundamental test charges.
This possibility seems to be experimentally ruled out, however, 
since no states of fractional baryon number have been observed.''}
Changing the gauge group from SU(3) to SU(2), it is interesting 
to note that, unlike quarks in QCD, the electron and neutrino in 
the electroweak Standard Model are not confined. 
The W$^{\pm}$ and Z$^0$ 
gauge bosons which couple to the neutrino 
are massive and the QED photon and QCD gluons are massless. 
What happens to the structure of non-perturbative propagators and vacuum energies when we turn off the coupling of the gauge bosons 
to left- or right-handed fermions?

Assuming the theory is ultraviolet consistent,
there are two issues to consider:
the pure SU(2) sector and also its coupling to QCD.
Pure Yang-Mills theory and Yang-Mills theory coupled to fermions are  both confining theories but the mechanism is different for each. 
Confinement is intimately connected with dynamical chiral symmetry  breaking
\cite{thooft,confinement}. 
Scalar confinement implies dynamical chiral symmetry breaking and a fermion condensate 
$\langle {\bar \psi} \psi \rangle < 0$.
For neutrinos, this is absent if there is no right-handed neutrino 
participating in the interaction.
Switching off the coupling of SU(2) gauge bosons to right-handed
fermions
must induce some modification of the non-perturbative propagators.
Either confinement is radically reorganised or
one goes to a Coulomb phase or to a Higgs phase
whereby the Coulomb force is replaced by a force of finite range with finite mass scale and the issues associated with infrared slavery are avoided. 
Additionally, going further,
QCD corrections dynamically break electroweak symmetry with 
Standard Model gauge interactions even with no Higgs condensate. 
The SU(2) gauge bosons couple to the quark axial-vector currents 
generating a small contribution to the mass of the SU(2) electroweak boson, 
about $g f_{\pi} \sim 30 \ {\rm MeV}$~\cite{weinstein} 
where $g$ is the SU(2) gauge coupling 
and $f_{\pi}$ is the pion decay constant.
This QCD correction vanishes if the QCD coupling is set to zero.

We next suppose the confinement to Higgs transition 
applies and explore possible consequences for particle physics.

Suppose that some process switches off 
the coupling of right-handed neutrinos to the SU(2) gauge fields.
In the electroweak Standard Model the electric charges of 
the quarks are fixed by the requirement of ultra-violet 
(axial-)anomaly cancellation in triangle diagrams involving 
three gauge boson legs when one sums over possible fermions 
in the triangle loop. 
Anomaly cancellation is required by gauge invariance and renormalisability.
If some dynamical process acts to switch off 
left- or right-handed fermions,
it will therefore have important consequences for the theory 
in the ultraviolet limit and should therefore be active there.
If symmetry breaking is dynamical and hence non-perturbative 
it will appear with coefficients smaller than any power of the running coupling. 
Following Ref.\cite{witten} we suppose an exponentially small effect.
Dynamical symmetry breaking then naturally induces a symmetry breaking scale $\Lambda_{\rm ew}$ 
which is much smaller than the high energy scales in the problem 
$M_{\rm cutoff}$
(which can be close to the Planck scale).
If we take the mass scale $M_{\rm cutoff}$ to be very large,
then the expression
\begin{equation}
\Lambda_{\rm ew}
 = M_{\rm cutoff} \ e^{-c/g(M_{\rm cutoff}^2)^2} 
\ \ll \ M_{\rm cutoff}
\end{equation}
naturally leads to hierarchies.
For example, the ratio of the weak scale $\Lambda_{\rm ew}$ 
to Planck mass is
$\Lambda_{\rm ew}/M_{\rm Pl} \sim 10^{-17}$.
For the mass scale in Eq.(6), $\Lambda_{\rm ew}/M \sim 10^{-14}$.
If symmetry breaking effects at very large scales are suppressed 
by the exponential 
$e^{-c/g(M_{\rm cutoff}^2)^2}$,
then 
$\Lambda_{\rm ew}$ 
is the mass scale appearing in the particle physics Lagrangian
describing the energy domain relevant to practical experiments.

\section{Spin model dynamics}

To help understand the different physics, 
we next consider a phenomenological trick 
to parametrise the different scales in the problem.

Analogies between quantum field theories and condensed matter and statistical systems have often played an important role in 
motivating ideas in particle physics.
Here we consider a possible analogy between the neutrino vacuum 
and the Ising model of statistical mechanics where the ``spins'' 
in the Ising model are associated with neutrino chiralities.

The ground state of the Ising model exhibits 
spontaneous magnetisation where all the spins line up; 
the internal energy per spin and the free energy density of the 
spin system go to zero. 
For an Ising system with no external magnetic field
the free energy density is equal to minus the pressure
\begin{equation}
P = - \biggl( {\partial F \over \partial V} \biggr)_{T}
\end{equation}
-- that is, the model equation of state looks like a 
vacuum energy term in 
Einstein's equations of General Relativity, $\propto g_{\mu \nu}$.

The Ising model uses a spin lattice to study ferromagnetism for a spin system in thermal equilibrium. 
One assigns a ``spin'' ($= \pm 1$) to each site and introduces a  nearest neighbour spin-spin interaction
\begin{equation}
H = 
- J \ \sum_{i,j} \ 
( \sigma_{i,j} \sigma_{i+1,j}  +  \sigma_{i,j+1} \sigma_{i,j} )
\ 
.
\end{equation}
Here $J$ is the bond energy and
we consider zero
external magnetic field.
Physical observables are calculated through the partition function 
$ Z = \sum_{\sigma_{ij=\pm1}} \exp ( - \beta H ) $
where $\beta = 1 / k T$, 
$k$ is Boltzmann's constant and $T$ is the temperature. 
One can normalise the energy by adding a constant so that neighbouring 
parallel spins give zero contribution. Then, the only positive  contribution to the energy will be from neighbouring disjoint spins of $2J$ and the probability for that will be $\exp (- 2 \beta J)$. 
Once a magnetisation direction is selected, it remains stable 
because  of the infinite number of degrees of freedom in the thermodynamic limit. 
The Ising model has a second order phase transition.
There is a critical coupling $(\beta J)_c$ so that for values of  $(\beta J) \geq (\beta J)_c$ 
the system develops a net magnetisation per spin ${\cal M} = \pm 1$,
the internal energy per spin and the free energy density each vanish
modulo 
corrections with 
the leading-term starting as a power of $\exp{(-2 \beta J)}$. 
The ground state ``vacuum" energy drops 
to a value close to zero from a very large value in the
phase transition which takes place close to the cut-off 
energy or temperature scale
$M_{\rm cutoff}$ and is induced by the ``spin" potential in the vacuum.

\subsection{Spin model neutrinos}

Can we construct a toy spin-model description for the neutrino vacuum?

First, the Ising-like interaction itself must be non-gauged otherwise  it will average to zero and there will be no spontaneous symmetry breaking and no spontaneous magnetisation \cite{creutz}.

Second, it is necessary to set a mass scale for $J$. 
If the spin model is to have connection with particle physics it is 
important to note that the coupling constant for the ``spin-spin'' 
interaction is proportional to the mass scale $J$. 
It therefore cannot correspond to a renormalisable interaction 
suggesting that fluctuations around the scale $J$ occur only near the  extreme high-energy limit of particle physics near the Planck mass. 
We consider the effect of taking $J \sim + M$. 
The combination $\beta J$ is then very large making it almost certain  that, if the spin model is applicable, the spontaneous magnetisation phase involving just left-handed neutrinos is the one relevant to  particle physics phenomena. 
The exponential suppression factor $e^{-2 \beta J}$ ensures that 
fluctuations associated with the Ising-like interaction are negligible in the ground state, 
thus preserving renormalisability for all practical purposes.

Setting the energy contribution of neighbouring parallel spins 
to zero in the Ising system is consistent here with 
zero net vacuum energy in particle physics with just 
left-handed neutrinos, normal ordering, 
no Higgs condensate and no QCD contribution.

Next, suppose we start with a gauge theory based on 
SU(3)$\otimes$SU(2)$\otimes$U(1)
coupled to quarks and leptons with no chiral dependent couplings, 
unbroken local gauge invariance and no elementary scalar Higgs field. 
(Here the SU(3) refers to QCD colour and SU(2)$\otimes$U(1) is the 
 electroweak gauge group.) 
We then turn on the spin model interaction coupled just 
to the neutrino in the upper component of 
the SU(2) isodoublet with the coupling 
$J \sim M \gg \alpha_s, \alpha_{\rm ew}, \alpha$ 
(the QCD, SU(2) weak and QED couplings). 
The gauge sector with small couplings acts like an ``impurity"
in the spin system.
It seems reasonable that the Ising interaction here exhibits 
the same two-phase picture with spontaneous magnetisation. 
Then, in the symmetric phase where $\beta J < (\beta J)_c$ 
the theory is symmetric under exchange of left and right handed neutrino chiralities and we have unbroken local gauge invariance. 
In the spontaneous magnetisation phase the neutrino vacuum is 
``spin''-polarised, a choice of chirality is made and the 
right-handed  neutrino decouples from the physics. 
Parity is spontaneously broken and the gauge theory coupled to the leptons becomes SU(2)$_{L} \otimes$U(1).
Following the discussion in Section 3, it seems reasonable to believe that the SU(2) gauge symmetry coupled to the neutrino is now spontaneously broken.

\subsection{Vacuum energy with spin model neutrinos}

Weak interactions mean that we have two basic scales in the problem: 
$J \sim M$ and the electroweak scale $\Lambda_{\rm ew}$ induced 
by spontaneous symmetry breaking.
For a spin model type interaction, the ground state with left-handed ``spin" chiralities is characterised by vanishing energy density.
Excitation of right-handed chiralities is associated with the large scale $2M$.
Then the mass scale associated with the vacuum for the ground state of the combined system (spin model plus gauge sector) 
one might couple to gravity reads in matrix form as 
\begin{equation} 
\mu_{\rm vac}
\sim 
\left[ \begin{array}{cc}
\! 0  &  - \Lambda_{\rm ew}  \! \\
\! - \Lambda_{\rm ew} &  -2M  \!
\end{array} \right] 
\end{equation}
with the different terms depending how deep we probe into the Dirac sea.
Here the first row and first column refer to left-handed states of the spin model ``neutrino" and the second row and second column refer to the right-handed states. 
The off-diagonal entries correspond to the potential in the 
vacuum associated with the dynamically generated Higgs sector.
Eq.(10) looks like the see-saw mechanism \cite{seesaw}.
Diagonalising the matrix for $M \gg \Lambda_{\rm ew}$ 
gives the light mass eigenvalue 
\begin{equation}
\mu_{\rm vac} 
\sim  \Lambda^2_{\rm ew} / 2M
\end{equation}
-- that is, the phenomenological result in Eq.(6).
Here 
the electroweak contribution $\Lambda_{\rm ew}$ 
is diluted by the ``spin'' potential in the vacuum.
The resultant picture is a 
Higgs sector characterised by 
scale 
$\Lambda_{\rm ew}$ embedded in the 
``spin" polarised ground state
that holds up to the ultraviolet scale $2M$.

In conclusion, the cosmological constant puzzle continues to fascinate.
Why is it finite, positive and so very small?
What suppresses the very large vacuum energy contributions expected from particle physics?
Understanding these vital questions will teach us much about 
the intersection of quantum field theory on the one hand, 
and gravitation on the other.

\section*{\bf Acknowledgements}

The research of SDB is supported by the Austrian Science Fund, 
FWF, through grants P20436 and P23753.
This work was completed at the 2012 Oberw\"olz Symposium
{\it Quantum Chromodynamics: History and Prospects}.
I thank H. Fritzsch and W. Plessas for the invitation 
to this stimulating meeting, celebrating 40 years of QCD.


\begin{thebibliography}{10}
%
\bibitem{weinberg}
S. Weinberg, Rev. Mod. Phys. 61 (1989) 1;
\\
V. Sahni and A. Starobinsky, Int. J. Mod. Phys. D 9 (2000) 373;
\\
S. M. Carroll, Living Rev. Relativity 3 (2001) 1;
\\
E. J. Copeland, M. Sami and S. Tsujikawa,
Int. J. Mod. Phys. D15 (2006) 1753;
\\
N. Straumann, Lect. Notes Phys. 721 (2007) 327;
\\
J. A. Frieman, M. S. Turner and D. Huterer,
Ann. Rev. Astron. Astrophys. 46 (2008) 385;
\\
R. Bean, {\tt arXiv:1003.4468 [astro-ph.CO]}.
%
\bibitem{astrid}
M-T. Jaekel, A. Lambrecht and S. Reynaud,
New Astron. Rev. 46 (2002) 727.
%
\bibitem{bass:2010}
S. D. Bass, J. Phys. G 38 (2011) 043201.
%
\bibitem{Ruiz}
{\it Dark energy: Observational and Theoretical Approaches},
(ed.) P. Ruiz-Lapuente (Cambridge UP 2010).
%
\bibitem{Amendola}
{\it Dark energy: Theory and Observations},
L. Amendola and S. Tsujukawa (Cambridge UP 2010).
%
\bibitem{bjd}
J. D. Bjorken and S. D. Drell,
{\it Relativistic Quantum Fields} (McGraw-Hill, 1965).
%
\bibitem{jaffe}
R. L. Jaffe, Phys. Rev. D72 (2005) 021301(R).
%
\bibitem{altarelli}
G. Altarelli,
Nucl. Phys. B (Proc. Suppl.) 143 (2005) 470,
{\tt hep-ph/0410101}.
%
\bibitem{wetterich}
R. Fardon, A. E. Nelson and N. Weiner, 
JCAP 10 (2004) 005; \\
C. Wetterich, Phys. Lett. B655 (2007) 201.
%
\bibitem{nucond}
D. G. Caldi and A. Chodos, {\tt hep-th/9903416}; \\
J. I. Kapusta, Phys. Rev. Lett. 93 (2004) 251801.
%
\bibitem{hfsola}
H. Fritzsch and J. Sola, 
Class. Quant. Grav. 29 (2012) 215002.
%%{\tt arXiv:1202.5097 [hep-ph]} 
%
\bibitem{gribov}
V. N. Gribov, Nucl. Phys. B206 (1982) 103.
%
\bibitem{gross}
D. J. Gross, I. R. Klebanov, A. V. Matytsin and A. V. Smilga,
Nucl. Phys. B461 (1996) 109.
%
\bibitem{thooft}
G. 't Hooft,
in {\it Recent developments in gauge theories},
ed. G. 't Hooft, Plenum, New York,
NATO Adv. Study Inst. Ser. B Phys. 59 (1980) 117.
%
\bibitem{confinement}
R. Alkofer, C. S. Fischer and F. J. Llanes-Estrada,
Mod. Phys. Lett. A23 (2008) 1105.
%
\bibitem{weinstein}
M. Weinstein, Phys. Rev. D8 (1973) 2511; \\
C. Quigg, Ann. Rev. Nucl. Part. Sci. 59 (2009) 505.
%
\bibitem{witten}
E. Witten, Nucl. Phys. B188 (1981) 513.
%
\bibitem{creutz}
M. Creutz, {\it Quarks, gluons and lattices} (Cambridge UP, 1983).
%
\bibitem{seesaw}
P. Minkowski, Phys. Lett. B67 (1977) 421;
\\
M. Gell-Mann, P. Ramond and R. Slansky, 
in {\it Supergravity}, edited
by D. Freedman and P. van Nieuwenhuizen (North Holland, Amsterdam, 1979),
p. 315;
\\
T. Yanagida,
in {\it Proceedings of the Workshop on Unified Theory and
Baryon Number in the Universe}, edited by O. Sawada and A. Sugamoto
(KEK, Tsukuba, Japan, 1979);
\\
R. Mohapatra and G. Senjanovic, Phys. Rev. Lett. 44 (1980) 912.
%

\end{thebibliography}
\end{document}